\documentclass[envcountsame,runningheads]{llncs}
\usepackage{a4}
\usepackage{times}
\usepackage{graphicx}
\usepackage{amssymb}
\usepackage{amsmath}
\usepackage{eucal}
\usepackage{pifont}
\newcommand{\ball}{B}
\newcommand{\cball}{\closure{\ball}}

\newcommand{\moo}{\operatorname{moo}}
\newcommand{\moc}{\operatorname{moc}}
\newcommand{\Moo}{\operatorname{Moo}}
\newcommand{\Moc}{\operatorname{Moc}}
\newcommand{\Time}{\operatorname{Time}}
\newcommand{\IQ}{{\mathbb{Q}}}
\newcommand{\IR}{{\mathbb{R}}}
\newcommand{\IN}{{\mathbb{N}}}

\newcommand{\IC}{{\mathbb{C}}}

\newcommand{\COMMENTED}[1]{}

\newcommand{\Open}{{\mathfrak{O}}}

\newcommand{\calO}{\mathcal{O}}
\newcommand{\Graph}{\operatorname{Graph}}
\newcommand{\Rel}[1]{\Graph(#1)}
\newcommand{\mycite}[2]{\cite[\textsc{#1}]{#2}}
\newcommand{\dom}{\operatorname{dom}}
\newcommand{\proj}{\operatorname{pr}}

\newcommand{\myrho}{\varrho}
\newcommand{\rhosd}{\myrho_{\textrm{sd}}}

\newcommand{\myto}{\!\to\!}
\newcommand{\toto}{\rightrightarrows}
\newcommand{\mapstoto}{\mapsto}
\newcommand{\person}[1]{\textsc{#1}}
\newcommand{\closure}[1]{\overline{#1}}
\newcommand{\interior}[1]{\overset{\circ}{#1}}
\newcommand{\myl}{{\scriptscriptstyle<}}
\newcommand{\myg}{{\scriptscriptstyle>}}
\newcommand{\myrhol}{\myrho_{\raisebox{0.2ex}{$\myl$}}}
\newcommand{\myrhog}{\myrho_{\raisebox{0.2ex}{$\myg$}}}
\newcommand{\psiL}[1]{\psi^{\hspace*{-0.7pt}#1}_{\!\raisebox{0.2ex}{$\myl$}}}
\newcommand{\psiG}[1]{\psi^{\hspace*{-0.7pt}#1}_{\!\raisebox{0.2ex}{$\myg$}}}

\newcommand{\kappaG}[1]{\kappa^{\hspace*{-0.7pt}#1}_{\!\raisebox{0.2ex}{$\myg$}}}

\newcommand{\mykappaG}{\kappaG{Y}}

\newcommand{\thetaL}[1]{\theta^{#1}_{\raisebox{0.5ex}{$\scriptscriptstyle<$}}}

\newcommand{\thetadl}{\thetaL{d}}

\newcommand{\thetal}{\theta_{\raisebox{0.5ex}{$\scriptscriptstyle<$}}}
\newcommand{\ithetaL}[1]{\overset{\circ}{\rule{0pt}{1.00ex}\smash{\thetaL{#1}}}}
\newcommand{\cthetaL}[1]{{\vphantom{\theta}\smash{\overline{\theta}}}^{\hspace*{-0.5pt}#1}_{\raisebox{0.5ex}{$\myl$}}}
\newcommand{\cthetadl}{\cthetaL{d}}
\newcommand{\IRc}{\IR_{\text{c}}}

\renewcommand{\vec}[1]{\mathbf{#1}}
\newcommand{\Ov}{\mathbf{0}}
\newcommand{\rank}{\operatorname{rank}}
\newtheorem{fact}[theorem]{Fact}
\newtheorem{myclaim}[theorem]{Claim}

\begin{document}
\title{Effectively Open Real Functions}
\author{Martin Ziegler}
\institute{IMADA, University of Southern Denmark, 5230 Odense M, DENMARK}
\maketitle
\begin{abstract}
A function $f$ is continuous iff
the \emph{pre}-image $f^{-1}[V]$ of any open set $V$
is open again. Dual to this topological property,
$f$ is called \emph{open} iff the \emph{image} $f[U]$
of any open set $U$ is open again. Several
classical Open Mapping Theorems in Analysis provide
a variety of sufficient conditions for openness.

By the Main Theorem of Recursive Analysis, computable
real functions are necessarily continuous.
In fact they admit a well-known characterization
in terms of the mapping $V\mapsto f^{-1}[V]$ being \emph{effective}:
Given a list of open rational balls exhausting $V$,
a Turing Machine can generate a corresponding list for $f^{-1}[V]$.
Analogously, \emph{effective openness} requires the mapping
$U\mapsto f[U]$ on open real subsets to be effective.

By effectivizing classical Open Mapping Theorems
as well as from application of Tarski's
Quantifier Elimination,
the present work reveals several rich classes of
functions to be effectively open.
\end{abstract}
\section{Introduction}
Computability theory over the reals started by investigating
single numbers \cite{Turing}. When real functions
were later considered it turned out that continuity was a
necessary condition for computability.
A function $f:X\to Y$ between topological spaces is continuous iff,
for any open set $V\subset Y$, its \emph{pre}-image
$f^{-1}[V]\subseteq X$ is open again.
In the case of open $X\subseteq\IR^n$ and $Y=\IR^m$
this means that, for any countable union of
$m$-dimensional open rational Euclidean balls
$$V=\;\bigcup\limits_j \ball(\vec y_j,r_j), \quad
  \vec y_j\in\IQ^m, \; r_j\in\IQ_{>0}, \qquad
  \ball(\vec y,r):=\{ \vec u\in\IR^m: |\vec y-\vec u|<r\} \enspace , $$
$U:=f^{-1}[V]\subseteq\IR^n$
is also a countable union of $n$-dimensional open rational Euclidean
balls $\ball(\vec x_\ell,s_\ell)$.
Moreover
$f$ is computable in the sense of \cite{Grzegorczyk,PER,Ko} 
iff the mapping $V\mapsto f^{-1}[V]$ on hyperspaces of open subsets
is effective in that, given a list of (centers $\vec x_k$ and
radii $r_k$ of) open rational Euclidean balls
$\ball(\vec x_k,r_k)\subseteq\IR^n$
exhausting $V,$ one can compute a corresponding list of
open rational Euclidean balls
$\ball(\vec y_\ell,s_\ell)\subseteq\IR^n$ exhausting $f^{-1}[V]$;
cf. \textsc{Lemma~6.1.7} in \cite{Weihrauch}.

So to speak `dual' to continuity is openness:
The function $f$ is \emph{open} if, rather than its pre-image,
its image $f[U]\subseteq Y$ is open
for any open set $U\subseteq X$.
While for example any constant $f$ lacks the latter
property, conditions
sufficient for its presence are given by a variety of
well-known \textsf{Open Mapping Theorems} for instance in
Functional Analysis,
Complex Calculus,
Real Analysis,
or Algebraic Topology.

The classical duality of continuity and openness raises the question
whether and to what extent it carries over to the computable setting.
For the first two aforementioned theorems, effectivized versions (in the
sense of Recursive Analysis) have been established respectively
in \cite{Banach} and \cite{Riemann}; see Theorem~\ref{thHertling} below.
It is indeed natural to consider, similarly to continuity and
computability, also \emph{effective} openness in the following sense:

\begin{definition} \label{defOpen}
Let $X\subseteq\IR^n$ be r.e.~open\footnote{that is, the union of
certain open rational balls $\ball(\vec z_j,t_j)$ whose
centers $\vec z_j$ and radii $t_j$ form 
computable rational sequences; cf.~\cite{Weihrauch}}.
Call an open function $f:X\to\IR^m$
\textsf{effectively open} if, from
any two lists
$(\vec x_j)_{_{j\in\IN}}\subseteq\IQ^n$
and $(r_j)_{_{j\in\IN}}\subseteq\IQ_{>0}$
such that $U=\bigcup_j \ball(\vec x_j,r_j)_{_j}\subseteq X$,
a Turing Machine can compute two similar lists
$(\vec y_\ell)_{_\ell}\subseteq\IQ^m$ and
$(s_\ell)_{_\ell}\subseteq\IQ_{>0}$
such that $f[U]=\bigcup_{\ell} \ball(\vec y_\ell,s_\ell)_{_\ell}$.
\end{definition}

In the convenient language of \emph{Type-2 Theory of Effectivity}
\cite{Weihrauch}, this amounts to the mapping
$U\mapsto f[U]$ on open Euclidean subsets
being $(\thetaL{n}\to\thetaL{m})$--computable.
Here, $\thetaL{d}$ denotes a canonical representation
for the hyperspace $\Open^d$ of open subsets of $\IR^d$;
cf.~\textsc{Definition~5.1.15} in \cite{Weihrauch}.

Apart from its natural duality to continuity and computability,
openness and effective openness arise
in the foundation of CAD/CAE \cite{Edalat}
in connection with regular sets --- i.e., 
roughly speaking, 
full-dimensional but not necessarily 
convex \cite{Kummer} ones --- 
as essential prerequisites for computations thereon;
cf. \textsc{Proposition~1.1}d-f) and
\textsc{Section~3.1} in \cite{MLQ2}.

The present work proves several rich and important
classes of functions to be effectively open
and thus applicable to such problems.
Our claims proceed in analogy to
those of classical Open Mapping Theorems.
An example due to \person{P.~Hertling} illustrates the idea:
\begin{theorem} \label{thHertling}
~\vspace*{-1ex}
\begin{enumerate}
\item
 Let $f:\IC\to\IC$ be complex differentiable and non-constant.
 Then $f$ is open.
\item
 Let $f$ furthermore be computable.
 Then it is effectively open.
\item
 Claim b) holds even uniformly in $f$,
 that is, the mapping ~$(f,U)\mapsto f[U]$~ with domain
 $$\big\{(f,U)\,\big|\,f:\IC\to\IC \text{ complex differentiable non-constant},
 U\subseteq\IC\text{ open}\big\}$$
 is
 $\big([\myrho^2\myto\myrho^2]\times\thetaL{2}\to\thetaL{2}\big)$--computable.
\end{enumerate}
\end{theorem}
\begin{proof}
a) is well-known in Complex Analysis; see, e.g.,
\cite[pp.231--233]{Rudin}.
For b) and c), cf. \textsc{Corollary~4.4} and \textsc{Theorem~4.3}
in \cite{Riemann}, respectively.
\qed\end{proof}
Here, $\myrho^2$ denotes the Cauchy representation
for the set $\IC$ of complex numbers, identified with $\IR^2$;
and $[\myrho^n\myto\myrho^m]$ is a natural representation
for continuous functions from $\IR^n$ to $\IR^m$;
see \textsc{Definitions~4.1.17} and \textsc{6.1.1} in \cite{Weihrauch}.

In the spirit of the above result,
we present in Section~\ref{secClassical}
several classical Open Mapping Theorems from Real Analysis and Algebraic
Topology; and in Sections~\ref{secEffective} and
\ref{secEffectivized} according effectivizations.
More precisely, proof-mining reveals
several classes of computable open
functions on Euclidean space to be effectively open.
We focus on claims similar to Theorem~\ref{thHertling}b),
that is, for fixed $f$ but uniformly in $U$.
Section~\ref{secTarski} takes a different approach
in devising `from scratch' proofs that computable open semi-algebraic
functions are effectively open; here, arguments are
based on Algebra and
\person{Tarski}'s \textsf{Quantifier Elimination}.
Section~\ref{secConverse} finally investigates the general
relation between computability and effective openness.
We conclude in Section~\ref{secConclusion} with a strengthening
of \mycite{Theorem~3.9}{MLQ2} based on the results
from Section~\ref{secEffectivized}.
\section{Classical Open Mapping Theorems} \label{secClassical}
We start with a characterization of open functions
resembling that of continuous ones:
\begin{lemma} \label{lemModulus}
Let $X\subseteq\IR^n$ be open ~ and denote
~$\cball(\vec x,s):=
  \{\vec v\in\IR^n:|\vec v-\vec x|\leq s\}$.%
\begin{enumerate}
\itemsep3pt
\item[a\,i)]
  A function $f:X\to\IR^m$ is continuous ~iff~ the mapping
  \begin{equation} \label{eqMoc1}
   \Moc_f:X\times\IN\to\IR, \quad
     (\vec x,k)\mapsto \sup\big\{s\geq0:
    f\big[\,\cball(\vec x,s)\cap X\big]
    \,\subseteq\, \ball\big(f(\vec x),2^{-k}\big)\big\}
   \end{equation}
  is strictly positive;
\item[a\,ii)] equivalently:
  to any $(\vec x,k)\in X\times\IN$,
  there exists an $\ell\in\IN$ such that
  \begin{equation} \label{eqMoc2}
    f\big[\ball(\vec x,2^{-\ell})\cap X\big]
    \;\subseteq\; \ball\big(f(\vec x),2^{-k}\big) \enspace .
  \end{equation}
\item[b\,i)]
  A function $f:X\to\IR^m$ is open ~iff~ the mapping
  \begin{equation} \label{eqMoo1}
   \Moo_f:X\times\IN\to\IR, \quad
     (\vec x,k)\mapsto \sup\big\{s\geq0:
    \cball\big(f(\vec x),s\big)
    \,\subseteq\, f\big[\ball(\vec x,2^{-k})\cap X\big]\big\}
   \end{equation}
  is strictly positive;
\item[b\,ii)] equivalently:
  to any $(\vec x,k)\in X\times\IN$,
  there exists an $\ell\in\IN$ such that
  \begin{equation} \label{eqMoo2}
    \ball\big(f(\vec x),2^{-\ell}\big)
    \;\subseteq\; f\big[\ball(\vec x,2^{-k})\cap X\big] \enspace .
  \end{equation}
\end{enumerate}
\end{lemma}
Both the function $\Moc_f$ according to Equation~(\ref{eqMoc1})
as well as any mapping
$\moc:X\times\IN\to\IN$ satisfying Equation~(\ref{eqMoc2})
for $\ell=\moc(\vec x,k)$
are known as \emph{the} or \emph{a}, respectively, (local)
\emph{modulus of continuity} of $f$;
cf., e.g., \cite{Encyclopaedia} or
\mycite{Definition~6.2.6}{Weihrauch}.
The apparent similarity suggests the following
\begin{definition} \label{defModulus}
$\Moo_f$ according to Equation~(\ref{eqMoo1}) is
\emph{the modulus of openness} of $f$;
call some mapping $\moo:X\times\IN\to\IN$ 
\emph{a modulus of openness} of $f$
if  Equation~(\ref{eqMoo2}) holds for $\ell=\moo(\vec x,k)$.
\end{definition}
It is not sufficient for a modulus
of openness do not suffice to be positive or defined
on a dense subset only:
\begin{example} \label{exDense}
$f:\IR\ni x\mapsto |x-\pi|$~ lacks openness
but $\Moo_f:\IQ\times\IN\to\IR$ is positive.
\end{example}
\begin{proof}[Lemma~\ref{lemModulus}]
\vspace*{-1ex}
\begin{enumerate}
\item[a\,i)]
 Let $\Moc_f$ be strictly positive and
 and $V\subseteq\IR^m$ open. To show that $f^{-1}[V]$ is
 open again, let $\vec x\in f^{-1}[V]$ be arbitrary.
 As $\vec y:=f(\vec x)\in V$ and $V$ is open,
 $\ball(\vec y,2^{-k})\subseteq V$ for some $k\in\IN$.
 Then for $s:=\Moc_f(\vec x,k)/2$,
 the open set $U:=\ball(\vec x,s)\cap X$ satisfies
 $$ \vec x\;\in\;U\;\subseteq\; f^{-1}\big[f[U]\big]
   \;\overset{(\ref{eqMoc1})}{\subseteq}\;
   f^{-1}\big[\ball\big(\vec y,2^{-k}\big)\big]
   \;\subseteq\; f^{-1}[V] $$
 that is, an entire open ball around $\vec x$ lying within $f^{-1}[V]$.
\\
 Conversely let $f$ be continuous, $\vec x\in X$ and $k\in\IN$.
 Therefore the pre-image $U:=f^{-1}[V]$ of
 $V:=\ball\big(f(\vec x),2^{-k}\big)$ is open
 and contains $\vec x$. In particular
 $\cball(\vec x,s)\subseteq U$ for some $s>0$
 and $\Moc_f(\vec x,k)\geq s$ is positive.
\item[a\,ii)]
 If $s:=\Moc_f(\vec x,k)>0$, then let
 $\ell:=\lfloor\log_2({1}/{s})\rfloor$;
 conversely, (\ref{eqMoc2}) yields $s:=2^{-\ell-1}$
 as a positive lower bound to $\Moc_f(\vec x,k)$.
\item[b\,i)]
  If $f$ is open, then its image $f[U]$ of the open set
  $U:=\ball(\vec x,2^{-k})\cap X\not=\emptyset$ is open again
  and thus contains, around the point $f(\vec x)\in f[U]$,
  some non-empty ball $\cball\big(f(\vec x),s\big)$ entirely;
  hence $\Moo_f(\vec x,k)\geq s>0$.
  \\
  Conversely let $U$ denote an open subset of $X$.
  To any $\vec y\in f[U]$, consider $\vec x\in U$ with $\vec y=f(\vec x)$
  and $k\in\IN$ such that $\ball(\vec x,2^{-k})\subseteq U$. Then
  $s:=\Moo_f(\vec x,k)/2$ satisfies
  $$\ball(\vec y,s)
   \;\overset{(\ref{eqMoo1})}{\subseteq}\;
  f[\ball(\vec x,2^{-k})\cap X]
  \;\subseteq\; f[U] \enspace . $$
  Therefore $f[U]$ is open.
\item[b\,ii)]
  Follows as in a\,ii). In particular it holds that
  ~$ f[U] \,=\, \bigcup_{\vec x\in U}
  \ball\big(f(\vec x),2^{-\moo(\vec x,k_{\vec x})}\big)$~
  for open $U\subseteq X$ whenever
  $k_{\vec x}\in\IN$ satisfies $\ball(\vec x,2^{-k_{\vec x}})\subseteq U$.
\qed
\end{enumerate}
\end{proof}

Many famous classical theorems give sufficient conditions
for a real function to be open. Several such claims are
collected in the following
\begin{fact} \label{facOpen}
Let $X\subseteq\IR^n$ be open.
\vspace*{-1ex}
\begin{enumerate}
\item
  Suppose continuous $f:X\to\IR$ has no local extrema (i.e.,
  to any open $U\subseteq X$ and $\vec x\in U$, there exist
  $\vec x_-,\vec x_+\in U$ such that
  $f(\vec x_-)<f(\vec x)<f(\vec x_+)$);
  then $f$ is open.
\item
  Any affinely linear mapping
  $\IR^n\ni\vec x\mapsto A\cdot\vec x+\vec b\in\IR^m$
  is open iff it is surjective.
\item
  Any continuously differentiable ('$C^1$')
  $f:X\to\IR^m$ is open, provided its
  \textsf{Jacobian}
  ~$f'(\vec x)=\big((\partial_i f_j)_{ij}\big)(\vec x)$~
  has rank ~$m$~ for all $\vec x\in X$.
\item
  Whenever continuous $f:X\to\IR^n$
  satisfies \emph{local injectivity} (i.e.,
  to each $\vec x\in X$ there exists $\varepsilon>0$ such that
  the restriction
  $f\big|_{\ball(\vec x,\varepsilon)}$ is injective),
  then it is open.
\end{enumerate}
\end{fact}
Claim~d) generalizes \textsf{Domain Invariance}
from {Algebraic Topology} where often 
injectivity is presumed globally.
\begin{proof}[Fact~\ref{facOpen}]
\vspace*{-1ex}
\begin{enumerate}
\item[a)]
  Exploit the one-dimensional range and apply
  the \textsf{Intermediate Value Theorem}:
  For an open ball $U:=\ball(\vec x,2^{-k})\subseteq X$,
  $f[U]$ is connected and thus a real interval.
  As furthermore $f$ has by prerequisite no local extrema,
  any $y\in f[U]$ is accompanied by $y_-,y_+\in f[U]$
  such that $y_-<y<y_+$. This implies
  $(y_-,y_+)\subseteq f[U]$ 
  and reveals that $f[U]$ is open.
\item[b)]
  follows from c), as the Jacobian of $f(\vec x)=A\cdot\vec x+\vec b$
  is $A\in\IR^{m\times n}$ (independent of $\vec x$) and $\rank(A)=m$
  is equivalent to $f$ being surjective.
\item[d)]
  See for example \mycite{Theorem~4.3}{Deimling}
  where (for $r=\varepsilon$) the proof proceeds by showing that
  the \emph{topological degree} $d(\Omega,f,\vec y)$ of $f$
  with respect to domain $\Omega:=\ball(\vec x,r)$
  is non-zero for all $\vec y$ in some $s$-ball around $f(\vec x)$.
  This guarantees that $f\big|_{\Omega}$ attains any such value
  $\vec y\in\ball\big(f(\vec x),s\big)$, that is,
  $f[\Omega]$ contains $\ball\big(f(\vec x),s\big)$.
  For $\varepsilon<2^{-k}$, this implies
  $\Moo_f(\vec x,k)\geq s/2>0$ and by 
  Lemma~\ref{lemModulus}b) yields openness of $f$.
\qed
\end{enumerate}
\end{proof}
Regarding c), $f$ has no chance of being locally injective whenever $n>m$
so that d) is not applicable in that case.
Instead, exploiting differentiability,
recall the \textsf{Inverse Function Theorem} from {Real Analysis}:
\begin{fact} \label{facInverse}
Let $U\subseteq\IR^n$ be open,
$f:U\to\IR^m$ continuously differentiable,
and $\vec x_0\in U$ such that $\rank f'(\vec x_0)=m$.
Then there exists a continuously differentiable local right
inverse to $f$ at $\vec x_0$, that is, $\delta>0$ and a $C^1$ function
\begin{equation} \label{eqInverse} \begin{gathered}
g:\ball\big(f(\vec x_0),\delta\big)\subseteq\IR^m
\to U\qquad\text{ such that} \\
g\big(f(\vec x_0)\big)=\vec x_0, \qquad
f\big(g(\vec y)\big)=\vec y \quad \forall
\vec y\in\ball\big(f(\vec x_0),\delta\big)
\end{gathered} \end{equation}
If $n=m$, then $g$ is unique and locally left inverse to $f$,
i.e.,  $g\big(f(\vec x)\big)=\vec x$
on $\ball(\vec x_0,\varepsilon)$ for some $\varepsilon>0$.
\end{fact}
In particular, $f[U]$ covers the open ball
$\ball\big(f(\vec x_0),\delta\big)\subseteq\IR^m$.
By taking $U:=\ball(\vec x_0,2^{-k})\subseteq X$,
we obtain $\Moo_f(\vec x,k)\geq\delta/2>0$ and
Fact~\ref{facOpen}c) finally follows with Lemma~\ref{lemModulus}b).
\section{Effective Continuity, Effective Openness} \label{secEffective}
The present section is about an effectivization of Lemma~\ref{lemModulus}.
While positivity of $\Moc_f$/$\Moo_f$ is trivially equivalent
to the existence of an according $\moc$/$\moo$, respectively,
similar equivalences are by no means obvious with respect
to computability. In fact for this purpose, both $\moc$ and $\moo$
have to be allowed to become \emph{multi-valued} in the sense 
\mycite{Definition~3.1.3.4}{Weihrauch}
that the integer $\ell$ returned by a Type-2 Machine computing
$\moc(\vec x,k)$ or $\moo(\vec x,k)$
may depend, rather than on the value of the argument $\vec x$ itself,
also on the particular choice of rational approximations for $\vec x$.
Such effects are well-known in Recursive Analysis,
see for instance
\cite[\textsc{Example~4.1.10} or \textsc{Theorem~6.3.7}]{Weihrauch}.

Also recall, e.g. from \cite{Weihrauch}, that $\myrhol$ is a
representation for $\IR$ connected to lower computability
in that it encodes rational approximations to the real
number under consideration \emph{from below}. Furthermore,
$\nu$ denotes the standard notation of $\IN$.
\pagebreak[2]%
\begin{theorem} \label{thModulus}
Let $X\subseteq\IR^n$ be r.e. open. 
Parallel to (the numbering in) Lemma~\ref{lemModulus}, we have:%
\vspace*{-1ex}%
\begin{enumerate}
\item[a\,ii)]
Fix some effective (i.e., $(\nu\to\myrho^n)$--computable)
enumeration $(\vec x_j)_{_j}$ of a dense subset\footnote{like,
for instance, $X\cap\IQ$} of $X$.
  A function $f:X\to\IR^m$ is computable iff
  the real sequence $\big(f(\vec x_j)\big)_j$ is computable
  and
  $f$ admits a $(\myrho^n\times\nu\toto\nu)$--computable
  multi-valued function
  $\moc:X\times\IN\toto\IN$
  such that Equation~(\ref{eqMoc2}) holds for all
  $\ell\in\moc(\vec x,k)$, ~$\vec x\in X$, $k\in\IN$.
\item[b\,i)]
  A computable $f:X\to\IR^m$ is effectively open iff
  $\Moo_f:X\times\IN\to\IR$ is strictly positive \quad and
  $(\myrho^n\times\nu\to\myrhol)$--computable;
\item[b\,ii)] equivalently: ~
  $f$ admits a $(\myrho^n\times\nu\toto\nu)$--computable
  {multi-valued} function
  $\moo:X\times\IN\toto\IN$ such that
  Equation~(\ref{eqMoo2}) holds for all
  $\ell\in\moo(\vec x,k)$, ~$\vec x\in X$, $k\in\IN$.
\end{enumerate}
\end{theorem}
Claim~a\,ii) is closely related to \textsc{Theorem~6} in \cite{Grzegorczyk}.
Extending Definition~\ref{defModulus},
multi-valued functions $\moc$/$\moc$
in the sense of Claims~a\,ii) and b\,ii) will in the sequel also be
called \emph{moduli of continuity/openness}, respectively.
Before turning to the proof, we provide in
Lemma~\ref{lemMulti} some tools on multi-valued
computability which turn out to be useful.

By the Main Theorem of Computable Analysis,
any computable real function $f$ on a compact
domain is continuous and thus bounded.
However the present work also considers
\emph{multi}-valued functions like moduli
of continuity; and such functions can in
general be unbounded even on compact domains.
\begin{example}
For a rational sequence $(x_j)_{_j}$ with
$|x_j-x|<2^{-j}$ for all $j=0,1,\ldots$, let
$$ F\big((x_j)_{_j}\big) \quad:=\quad
   \lfloor \tfrac{1}{x_0+1} \rfloor \enspace . $$
Then, $F$ is a computable realization of a
multi-valued, unbounded function $f:[0,1]\toto\IN$.
\end{example}
Item~b) below basically says that such unpleasant
cases can always be avoided by passing to another
computable multi-valued function. To this end, we
call $\tilde f:X\toto Y$ a \emph{sub-function} of $f:X\toto Y$
if $\tilde f(\vec x)\subseteq f(\vec x)$ for all $\vec x\in X$
and remark that, according to
\mycite{Definition~3.1.3.4}{Weihrauch},
if $\tilde f$ is computable then so are
\emph{all} its super-functions $f$
\begin{lemma} \label{lemMulti}
Let $X\subseteq\IR^n$ be r.e. open and let
$\rhosd$ denote the \emph{signed digit} representation.
\vspace*{-1ex}
\begin{enumerate}
\item
  The multi-valued partial mapping
  $G:\subseteq\Open^n\times\IR^n\toto\IN$,
  $\Rel{G}:=\big\{(U,\vec x,k):\cball(\vec x,2^{-k})\subseteq U\big\}$
  ~ is $(\thetaL{n}\times\myrho^n\toto\nu)$--computable.
\item
  To every $(\myrho^n\toto\myrho^m)$--computable
  multi-valued $f:X\toto\IR^m$,
  there exists a multi-valued
  $(\myrho^n\toto\myrho^m)$--computable
  sub-function $\tilde f$ for which the image 
  \smash{$\tilde f[K]:=\bigcup\limits_{\vec x\in K} \tilde f(\vec x)$}
  $\subseteq\IR^m$
  of any compact subset $K$ of $X$ is bounded.
\item
  For $m=1$ and the function $\tilde f$ from b),
  an upper bound $N\in\IN$ on $\tilde f[K]\subseteq\IR$ can be found
  effectively; formally:
  The multi-valued mapping $K\mapstoto N$ with
  $\tilde f[K]\subseteq [-N,+N]$ is
  $(\kappaG{n}\toto\nu)$--computable.
\end{enumerate}
Claims b) and c) also hold uniformly in $p$ for parametrized
computable functions $p\mapsto f(p,\,\cdot\,):X\toto\IR^m$.
\end{lemma}
\begin{proof}
\begin{enumerate}
\item
  By \cite[\textsc{Lemma~4.1}b)]{MLQ2}, the property
  ~``$\cball(\vec x,2^{-k})\subseteq U$''~ is semi-decidable
  with respect to $k\in\IN$ for $\vec x$ given by a
  $\myrho^n$--name and $U\in\Open^n$ by a
  $\thetaL{n}$--name. So dove-tailed search can effectively
  find an appropriate $k$ whenever $\vec x\in U$.
\item
  Let $F:\subseteq\Sigma^\omega\to\Sigma^\omega$ denote some
  computable (single-valued) realization of $f$.
  Exploiting $\myrho\equiv\rhosd$
  according to \mycite{Theorem~7.2.5.1}{Weihrauch},
  we pre-compose $F$ with a computable function
  $G$ converting
  $\rhosd^n$-names to $\myrho^n$-names.
  $F\circ G$ therefore realizes a
  $(\rhosd^n\toto\myrho^m)$--computable sub-function
  $\tilde f$ of $f$, defined by
  $\tilde f(\vec x)=\big\{ \myrho^m(\bar\sigma) :
    \rhosd^n(\bar\sigma)=\vec x\big\}$.
  By \mycite{Exercise~7.2.9}{Weihrauch},
  the collection $\tilde K\subseteq(\Sigma^\omega)^n$
  of all $\rhosd^n$-names $\bar\sigma$ of
  all $\vec x\in K$ is in particular compact.
  Being Cantor-continuous, $F\circ G$ maps $\tilde K$
  to a compact set $(F\circ G)[\tilde K]$ whose
  image under $\myrho^m$, namely the set $\tilde f[K]$,
  is again compact by admissibility of $\myrho^m$.
\item
  Rather than carefully adapting the proof of
  for example \mycite{Theorems~7.1.5}{Weihrauch},
  we slightly modify the Type-2 Machine $M$
  computing $F\circ G$ in b) to operate as follows:
  Upon input of a $\rhosd^n$-name for $\vec x\in X$
  and while calculating rational approximations
  $y_j$ to $y=\tilde f(\vec x)$ with $|y_j-y|<2^{-j}$,
  idly loop $\lceil |y_0|+1\rceil$
  times before actually outputting the first symbol
  of that $\myrho$-name for $y$ and then proceeding
  like $M$.

  This new machine $\tilde M$ will thus satisfy
  $\dom(M)=\dom(\tilde M)$ and
  $\Time_{\tilde M}(\bar\sigma)(1)\geq\tilde f(\vec x)$
  for any $\rhosd^n$-name $\bar\sigma$ of $\vec x\in X$.
  In particular, $\Time_{\tilde M}^{\tilde K}(1)\in\IN$
  is an upper bound on $\tilde f[K]$
  where $\tilde K\subseteq\Sigma^\omega$ denotes the
  collection of all $\rhosd^n$-names $\tilde f$ for all
  $\vec x\in K$.

  According to \mycite{Exercise~7.2.9}{Weihrauch},
  $K\mapsto\tilde K$ is $(\kappaG{n}\to\mykappaG)$--computable;
  and \cite[\textsc{Exercise~7.1.4}a)]{Weihrauch} implies
  that, from a $\mykappaG$--name of $\tilde K$,
  one can effectively obtain an upper bound
  $N$ on $\Time_{\tilde M}^{\tilde K}(1)$.
\qed
\end{enumerate}
\end{proof}

\subsection{Proof of Theorem~\ref{thModulus}}
This section proves the several claims made
in Theorem~\ref{thModulus}.
\begin{claim}
Let $X\subseteq\IR^n$ be r.e. open,
$(\vec x_j)_{_j}$ a computable sequence dense in $X$,
and $f:X\to\IR^m$ computable.
Then the sequence $\big(f(\vec x_j)\big)$ is computable,
and $f$ admits a computable multi-valued
modulus of continuity.
\end{claim}
\begin{proof}
The first sub-claim is immediate. For the second one,
let $\vec x\in X$ and $k\in\IN$ be given.
From these, $\thetadl$--compute
$U:=f^{-1}\big[B\big(f(\vec x),2^{-k}\big)\big]\cap X$
by virtue of
\cite[\textsc{Theorem~6.2.4.1} and \textsc{Corollary~5.1.18.1}]{Weihrauch}.
Then invoke Lemma~\ref{lemMulti}a)
to obtain some $\ell\in G(U,\vec x)$.
This satisfies Equation~(\ref{eqMoc2}) because
$f[V]\subseteq U$ is equivalent to $V\subseteq f^{-1}[U]$.
\qed\end{proof}

\begin{claim}
Let $X\subseteq\IR^n$ be r.e. open,
$(\vec z_j)_{_j}$ a computable sequence dense in $X$,
$f:X\to\IR^m$ such that $\big(f(\vec z_j)\big)$ is computable,
and $\moc:X\times\IN\toto\IN$ a computable multi-valued
modulus of continuity. Then, $f$ is computable.
\end{claim}
\begin{proof}
First note that $f$ is continuous by Lemma~\ref{lemModulus}a).
We show that it furthermore admits effective evaluation:
Given a sequence $\vec x_\ell\in X$ of rational vectors with
$|\vec x-\vec x_\ell|<2^{-\ell}$ for some $\vec x\in X$,
one can computably obtain a sequence
$\vec y_k$ such that
$|f(\vec x)-\vec y_k|<2^{-k}$.

Indeed, calculate by prerequisite $\ell\in\moc(\vec x,k)$;
then search (dove-tailing) for some $j$ with
$\vec z_j\in X$ and
$|\vec z_j-\vec x_{\ell+1}|<2^{-\ell-1}$;
finally let $\vec y_k:=f(\vec z_j)$.
It follows $|\vec z_j-\vec x|<2^{-\ell}$
and thus, by Equation~(\ref{eqMoc2}),
$|f(\vec x)-\vec y_k|<2^{-k}$.
\qed\end{proof}

\begin{claim}
Let $X\subseteq\IR^n$ be r.e. open,
$f:X\to\IR^m$ computable and effectively open.
Then, $\Moo_f:X\times\IN\to\IR$ is
$(\myrho^n\times\nu\to\myrhol)$--computable.
\end{claim}
\begin{proof}
The mapping $(\vec x,k)\mapsto\ball(\vec x,2^{-k})\cap X=:U$
is $(\myrho^n\times\nu\to\thetaL{n})$--computable since
$X$ is r.e. open \mycite{Corollary~5.1.18.1}{Weihrauch}.
By assumption on effective openness of $f$,
one can therefore obtain a $\thetaL{m}$-name
for the open set $V:=f[U]\ni\vec y:=f(\vec x)$.
Then searching all rational $s\geq0$ satisfying
~$\cball\big(f(\vec x),s\big)\subseteq V$~
is possible due to \cite[\textsc{Lemma~4.1}b)]{MLQ2}
and yields lower approximations to (i.e., a
$\myrhol$--name for) the value $\Moo_f(\vec x,k)$.
\qed\end{proof}

\begin{claim}
Let $\Moo_f:X\times\IN\to\IR$ be strictly positive and
$(\myrho^n\times\nu\to\myrhol)$--computable;
then there is a computable multi-valued $\moo$.
\end{claim}
\begin{proof}
From a $\myrhol$--name of $s:=\Moo_f(\vec x,k)>0$,
obtain some $\ell\in\IN$ with $2^{-\ell}<s$;
compare \mycite{Example~4.1.10}{Weihrauch}.
\qed\end{proof}

For the converse claims in Theorem~\ref{thModulus}b),
the prerequisite of a computable $f$ can actually be
relaxed to continuity with computable values on
a computable dense subset. This resembles conditions
(9a) and (9b) in \cite{Grzegorczyk} and is, without
(9c) therein, more general than requiring computability of $f$.
\begin{myclaim} \label{clTwo}
Let $X\subseteq\IR^n$ be r.e. open,
$(\vec x_j)_{_j}$ a dense computable sequence in $X$,
$f:X\to\IR^m$ continuous,
the sequence $\big(f(\vec x_j)\big)_j$ computable,
and $\moo$ a computable multi-valued modulus of openness.
Then $f$ is effectively open.
\end{myclaim}
\begin{proof}
From Lemma~\ref{lemModulus}b) we already
know that $f$ is open. The goal is thus to
$\thetaL{m}$-compute $f[U]$, given a $\thetaL{n}$-name of some
open $U\subseteq X$.
The proof of Lemma~\ref{lemModulus}b\,ii) has revealed that
\begin{equation} \label{eqUnion}
f[U] \;\;\overset{\surd}{=}\;
 \bigcup_{\vec x\in U}
  \ball\Big(f(\vec x),2^{-\ell_{\vec x}}\Big)
\;\;\supseteq\
 \bigcup_{j:\vec x_j\in U}
  \ball\Big(f(\vec x_j),2^{-\ell_{\vec x_j}}\Big)
\;=:\, V
\end{equation}
for arbitrary $\ell_{\vec x}\in\moo\big(\vec x,G(U,\vec x)\big)$ 
with $G$ from Lemma~\ref{lemMulti}a).
$V$ is indeed contained in $f[U]$ as the union to the right
ranges only over certain $\vec x\in U$
compared to all in the left one.
Being only a countable union,
$V$ can be $\thetaL{m}$-computed
according to \textsc{Example~5.1.19.1} in \cite{Weihrauch}.
More precisely, the $\thetaL{n}$-name of $U$ permits enumeration
of all $j$ such that $\vec x_j\in U$ by virtue of
Lemma~\ref{lemMulti}a); the multi-valued mapping
$h:(U,\vec x)\mapstoto\moo\big(\vec x,G(U,\vec x)\big)$
is $(\thetaL{n}\times\myrho^n\toto\nu)$--computable;
and the multi-valued mapping $j\mapstoto
  \ball\big(f(\vec x_j),2^{-h(U,\vec x_j)}\big)$
is $(\nu\toto\thetaL{m})$--computable since $j\mapsto\vec x_j$,
$j\mapsto f(\vec x_j)$ both are by assumption.

To complete the proof of Claim~\ref{clTwo}, we shall show
that in fact the reverse inclusion ``$f[U]\subseteq V$''
holds as well for any $\ell_{\vec x_j}\in\tilde h(U,\vec x_j)$ 
with $\tilde h$ denoting the computable sub-function
according to Lemma~\ref{lemMulti}b).
So take arbitrary $\vec y\in f[U]$, $\vec y=f(\vec x)$
with $\vec x\in U$. If $\vec x=\vec x_j$ for some $j$,
then $\vec y\in V$ by definition anyway.
If $\vec x$ does not occur within the sequence $(\vec x_j)_{_j}$,
consider some compact
ball $\cball:=\closure{\ball(\vec x,r)}$
sufficiently small to be contained in $U$.
By the parametrized version of Lemma~\ref{lemMulti}b),
there exists\footnote{Here we do not
need to find this bound effectively.} an upper bound $L\in\IN$
for $\tilde h(U,\cdot)$ on $\cball$. Exploiting continuity,
$|f(\vec x)-f(\vec z)|<2^{-L}$ for all
$\vec z$ sufficiently close to $\vec x$.
In particular for an appropriate $\vec z=\vec x_j$
and any $\ell\in\tilde h(U,\vec x_j)\leq L$ by choice of $L$,
it holds that
$f(\vec x)\in\ball\big(f(\vec x_j),2^{-L}\big)
\subseteq\ball\big(f(\vec x_j),2^{-\ell}\big)$.
The latter term occurs in the right
hand side union of (\ref{eqUnion});
we have thus proven an arbitrary
$\vec y=f(\vec x)\in f[U]$ to lie in $V$.
\qed\end{proof}
\section{Effectivized Open Mapping Theorems} \label{secEffectivized}
Here come the already
announced effectivizations of
the classical claims from Fact~\ref{facOpen}.

\begin{theorem} \label{thOpen}
Let $X\subseteq\IR^n$ be r.e. open.
\vspace*{-1ex}
\begin{enumerate}
\item
  Every computable open $f:X\to\IR$ (i.e., with one-dimensional
  range) is effectively open. 
  More generally whenever a computable open $f:X\to\IR^m$ maps
  open balls $B\subseteq X$ to convex sets $f[B]\subseteq\IR^m$,
  then it is effectively open.\COMMENTED{
  The same holds\footnote{We wonder whether any continuous open
  $f$ mapping balls to convex sets also maps simplices to convex
  sets and/or conversely.}
  if $f$ maps open simplices\footnote{An open simplex
  $S\subseteq\IR^n$ is the topological interior of the convex hull
  of $n+1$ points in $\IR^n$,
  ~ $S=\{\sum_{i=0}^n \lambda_i \vec x_i : \vec x_i\in\IR^n,
      0\leq\lambda_i, \sum\lambda_i=1\}^\circ$.}
  to convex sets.}
\item
  Any surjective computable affinely linear mapping
  is effectively open.
\item If computable $f:X\to\IR^m$ is $C^1$,
  $f'$ is computable and
  has rank ~$m$~ everywhere, then $f$ is effectively open.
\item
  Let both $f:X\to\IR^n$ and $h:X\toto\IN$ be computable
  such that, for any $\vec x\in X$ and $\ell\in h(\vec x)$,
  the restriction $f\big|_{\ball(\vec x,2^{-\ell})\cap X}$ is injective.
  Then $f$ is effectively open.
\item
  Suppose $X$ is bounded, $f:\closure{X}\to\IR^n$ computable and locally
  injective; then $f\big|_X$ is effectively open. \hfill
  The same holds if $f:X\to\IR^n$ is computable and globally injective.%
\end{enumerate}
\end{theorem}
\begin{proof}
Claims~a) and d) will be proven in Subsections~\ref{subsecOnedim}
and \ref{subsecDeimling} respectively.
Claim~b) follows from c) just like in the classical case;
and c) in turn, once again similarly to the classical case,
is a consequence of the
\emph{effectivized} Inverse Function Theorem~\ref{thInverse} below.

Claim~d) implies the second part of e) as,
$h(\vec x):\equiv0$ will do.
For the first part of e) observe that,
$f$ being locally injective on a \emph{compact} domain,
finitely many out of the balls
$\ball\big(\vec x,\varepsilon(\vec x)\big)$
suffice to cover $\closure{X}$. Therefore
there exists one $\varepsilon_0>0$ common to all $\vec x\in X$
such that $f\big|_{\ball(\vec x,\varepsilon_0)\cap X}$ is injective;
w.l.o.g. $\varepsilon_0=2^{-k_0}$ for some $k_0\in\IN$.
As $k_0$ is just an integral constant,
$h(\vec x):\equiv k_0$ defines a computable
function; now apply Item~d).
\qed
\end{proof}

The following is a computable counterpart to
Fact~\ref{facInverse}:
\begin{theorem}[Effectivized Inverse Function Theorem] \label{thInverse}
Let $U\subseteq\IR^n$ be r.e. open,
$f:U\to\IR^m$ computable with computable derivative,
and $\vec x_0\in U$ computable such that $\rank f'(\vec x_0)=m$.
Then there exists a computable local right inverse to $f$,
that is, a computable function $g$ with computable derivative
satisfying (\ref{eqInverse})
and $\dom(g)=\ball\big(f(\vec x_0),\delta\big)\subseteq\IR^m$
for some rational $\delta>0$.

Moreover, such a $\delta=2^{-k}>0$ is uniformly
computable from
$\vec x_0$; formally: \\
For r.e. open $X\subseteq\IR^n$ and computable,
continuously differentiable $f:X\to\IR^m$ with computable derivative
$f':X\to\IR^{m\times n}$, the multi-valued mapping
$I:\Open^n\times X\toto\IN$ with
\vspace*{-1ex}\begin{multline*}
\Rel{I} := \big\{
   (U,\vec x_0,k)\,\big|\,
     \vec x_0\in U\subseteq X,\, \rank\big(f'(\vec x_0)\big)=m,\,
     \; \\ \exists
       g:\ball\big(f(\vec x_0),2^{-k}\big)\to U
      \text{ satisfying (\ref{eqInverse})}\big\}
\end{multline*}
is $(\thetaL{n}\times\myrho^n\toto\nu)$--computable.
\end{theorem}
Similarly to the classical case,
$f[U]$ in particular covers the open ball
$\ball\big(f(\vec x_0),\delta\big)\subseteq\IR^m$.
Setting $\moo(\vec x,k):=I\big(\ball(\vec x,2^{-k}),\vec x\big)$
therefore proves Theorem~\ref{thOpen}c) by virtue of
Theorem~\ref{thModulus}b).

We emphasize that Theorem~\ref{thInverse} can be generalized to hold even
uniformly in $(f,f')$. Furthermore the multi-valued computation
is extendable to yield not only $\delta$ but also $g$ and $g'$.
As the domain of these partial functions varies, an according
formalization however requires an appropriate representation
such as $\delta_1$ from \textsc{Exercise~6.1.11} in \cite{Weihrauch}
and is beyond our present interest.
\\
Let us also point out that, although the proofs to Theorem~\ref{thInverse}
(in Subsection~\ref{proofInverse})
as well as the one to Theorem~\ref{thOpen}c+d)
proceed by presenting according algorithms,
they are not necessarily constructive in the
intuitionistic sense since
the correctness of these algorithms relies on
\person{Brouwer}'s \textsf{Fixed-Point Theorem}.
\subsection{Proof of Theorem~\ref{thOpen}a)}
\label{subsecOnedim}
\begin{claim}
Let $X\subseteq\IR^n$ be r.e. open. 
If computable open $f:X\to\IR^m$ maps
open Euclidean balls 
to convex sets, then it is effectively open.
\end{claim}
\begin{proof}
Recall that a $\thetaL{n}$-name for $U\subseteq X$ is (equivalent to)
a list of all closed rational Euclidean balls
$\cball_i=\closure{\ball(\vec z_i,r_i)}$ contained in $U$
\cite[\textsc{Definition~5.1.15.1} and \textsc{Exercise~5.1.7}]{Weihrauch}.

Since it is easy to obtain a $\psiL{n}$-name for each such $\cball_i$,
one can $\psiL{m}$-compute $\closure{f[\cball_i]}$
by virtue of \textsc{Theorem~6.2.4.3} in \cite{Weihrauch}.
In fact $\closure{f[\cball_i]}=f[\cball_i]=\closure{f[\ball_i]}$
since $f$ is continuous and $\cball$ compact;
cf. \cite[\textsc{Lemma~4.4}d)]{MLQ2}.
The prerequisite asserts $f[\ball_i]$ to be convex,
and its closure is thus convex and even regular by
\textsc{Proposition~1.1}f) in \cite{MLQ2}.
By virtue of \cite[\textsc{Theorem~4.12}a)]{MLQ},
the $\psiL{m}$-name for $\closure{f[\ball_i]}$
can hence be converted into a matching
$\ithetaL{m}$-name, that is, a $\thetaL{m}$-name for
$\interior{\closure{f[\ball_i]}}$,
a subset of $f[U]$ as $\cball_i\subseteq U$ and
$\closure{f[\ball_i]}=f[\cball_i]$.

Doing so for all $\cball_i$ listed in the $\thetaL{n}$-name
of $U$ and taking their countable union
according to \mycite{Exercise~5.1.19}{Weihrauch},
we thus obtain a $\thetaL{n}$-name for some open subset $V$
of $f[U]$. To see that $V$ in fact coincides with $f[U]$,
consider some $\vec y\in f[U]$, $\vec y=f(\vec x)$ with $\vec x\in U$.
Then some entire ball $\ball(\vec x,s)$ is contained inside of $U$.
By density, there exist $\vec z\in\IQ^n$ and $0<r\in\IQ$ such that
$\vec x\in\ball(\vec z,r)\subseteq\cball(\vec z,r)\subseteq\ball(\vec x,s)$.
This $\cball:=\cball(\vec z,r)$ will thus occur in the input
and, in the union constituting $V$, lead to a term
$\interior{\closure{f[\ball]}}\supseteq f[\ball]\ni f(\vec x)=\vec y$,
cf. \cite[\textsc{Lemma~4.2}i)]{MLQ2}.
As $\vec y\in f[U]$ was arbitrary, this proves $V\supseteq f[U]$.
\qed\end{proof}
\subsection{Proof of Theorem~\ref{thOpen}d)}
\label{subsecDeimling}
Regarding Theorem~\ref{thModulus}b),
the claim follows uniformly in $f$ from the below
\begin{lemma} \label{lemDeimling}
Fix r.e. open $X\subseteq\IR^n$. The multi-valued mapping
$H:C(X,\IR^n)\times X\times\IN\toto\IN$\!\!
\begin{multline*} 
\text{with ~ }
\Rel{H}\;:=\; \big\{
  (f,\vec x,k,\ell)\,\big|\,
  f:X\to\IR^n \text{ injective on }
    \ball(\vec x,2^{-k})\subseteq X,\: \\
    \ball\big(f(\vec x),2^{-\ell}\big)
    \subseteq f[\ball(\vec x,2^{-k})]
    \big\}
\end{multline*}
is $([\myrho^n\myto\myrho^n]\times\myrho^n\times\nu\toto\nu)$--computable.
\end{lemma}
\begin{proof}
The mapping's value is indeed an open set because of
Fact~\ref{facOpen}d). Recall its proof based on
\textsc{Theorem~4.3} in \cite{Deimling} together
with \textsc{Theorem~3.1}(d4+d5) therein.
The latter reveal that, for each $\Omega:=\ball(\vec x,2^{-k-1})$
--- observe $\closure{\Omega}\subseteq X$ ---
$f[\Omega]$ covers $\ball\big(f(\vec x),r\big)$
where $r>0$ denotes the distance of $f(\vec x)$ to
the set $K:=f\big[\partial\Omega\big]$.
The sphere boundary $\partial\Omega$ being obviously
$\kappa^n$--computable from $(\vec x,k)$,
$K$'s distance function is uniformly computable
by virtue of \textsc{Theorem~6.2.4.4} in \cite{Weihrauch}.
In particular, one can effectively evaluate this
function at $f(\vec x)$ and thus obtain the aforementioned $r$.
From this it is easy to get
some $\ell\in\IN$ with $2^{-\ell}<r$.
\qed\end{proof}
\subsection{Proof of Effectivized Inverse Function Theorem}
\label{proofInverse}
An important part in the proof of Theorem~\ref{thInverse}
relies on the following result
on computability of unique zeros of real functions.
It generalizes \textsc{Corollary~6.3.5} in \cite{Weihrauch}
from one to higher dimensions.
\begin{lemma} \label{lemZero}
Consider the class of continuous real functions $f$
in $n$ variables on the closed unit ball
$\cball(0,1)\subseteq\IR^n$
attaining the value zero in \emph{exactly} one point.
Hereon, the $\cball$-valued function $Z_u$, defined by
$$ Z_u(f)=\vec x  \quad :\Longleftrightarrow\quad
  \vec x\text{\rm\;\;is the (unique) zero of\;\;}f \enspace , $$
is $([\myrho^n\myto\myrho]\to\myrho^n)$--computable.
\end{lemma}
\begin{proof}
By \textsc{Theorems~6.2.4.2} and \textsc{5.1.13.2} in \cite{Weihrauch}
one can, given a $[\myrho^n\myto\myrho]$-name of $f$,
$\psiG{n}$-compute the set $f^{-1}[\{0\}]\subseteq\cball$.
This computation actually yields a $\kappaG{n}$-name
of this set which, by prerequisite, consists of exactly
one point. Now apply \textsc{Exercise~5.2.3} in \cite{Weihrauch}.
\qed\end{proof}

Recall the second claim from Theorem~\ref{thInverse}
which shall be proven first:
\begin{myclaim} \label{clInverseB}
Let $X\subseteq\IR^n$ be r.e. open,
$f:X\to\IR^m$ computable with computable derivative
$f':X\to\IR^{m\times n}$.
Then the multi-valued mapping
$I:\Open^n\times X\toto\IN$ with
\begin{multline*}
\Rel{I} := \Big\{
   (U,\vec x_0,\ell)\,\big|\,
     \vec x_0\in U\subseteq X,\, \rank\big(f'(\vec x_0)\big)=m,\,
     \; \\ \exists
       g:\ball\big(f(\vec x_0),2^{-\ell}\big)\to U
      \text{ satisfying (\ref{eqInverse})}\Big\}
\end{multline*}
is $(\thetaL{n}\times\myrho^n\toto\nu)$--computable.
\end{myclaim}
\begin{proof}
Given a $\thetaL{n}$-name of $U\subseteq X$ and $\vec x_0\in U$,
determine according to Lemma~\ref{lemMulti}a) some
$k_0\in\IN$ such that
$\ball(\vec x_0,2^{-k_0})\subseteq U$.
Exploit differentiability of $f$ to write
\begin{equation} \label{eqDifferential}
f(\vec x) \quad=\quad
  f(\vec x_0) \;+\; f'(\vec x_0)\cdot(\vec x-\vec x_0) \;+\; r(\vec x)
\end{equation}
with computable and computably differentiable
$r$ satisfying $r(\vec x)/|\vec x|\to0$ as $\vec x\to\vec x_0$.
\begin{itemize}
\item
Since the computable
matrix-valued function $\vec x\mapsto f'(\vec x)$ was required
to have rank $m$ in $\vec x_0$,
certain $m$ of its columns are linearly independent.
In fact, one can effectively find a regular
$m\times m$ submatrix $A=A(\vec x_0)$ of $f'(\vec x_0)$: by
dove-tailing w.r.t.
all (finitely many) possible candidates and looking for one with
non-zero determinant. \\
For ease of notation,
suppose that $f'(\vec x_0)$ is of the form
$(A|B)$ with $B\in\IR^{(n-m)\times m}$.
Continuity of the function
 $\vec x\mapsto\det A(\vec x)$
with non-zero value at $\vec x_0$
yields that $A(\vec x)$ is regular
on a whole ball around $\vec x_0$;
$\vec x\mapsto\det A(\vec x)$ even being computable, a corresponding radius
$2^{-k_1}\leq2^{-k_0}$ can in fact be found effectively.
\item
By (computable) translation,
it suffices
to prove the claim for the computable function on only $m$ variables
$$\tilde f\;:\;\IR^m\supseteq\ball(\Ov,2^{-k_0})
\quad\ni\quad\vec x\quad\mapsto\quad
f(\vec x_0+\vec x)\quad\in\quad\IR^m \enspace . $$
Indeed, any local right inverse
$\tilde g:\ball\big(\tilde f(\vec x_0),\tilde\delta\big)\subseteq\IR^m\to
\tilde U:=\ball(\Ov,2^{-k_0})\subseteq\IR^m$
for this restriction can straight-forwardly (and computably)
be extended to one for $f$ by letting
$g(\vec y):=\big(\tilde g(\vec y),\Ov\big)+\vec x_0\in\IR^n$.
\item
$A=\tilde f'(\Ov)$
being regular, $c:=\min_{|\vec x|=1} |A\cdot\vec x|$
is non-zero and,
according to \textsc{Corollary~6.2.5} in \cite{Weihrauch},
can be effectively calculated from the given data.
\item
Effective continuity of $r'(\,\cdot\,)$ together with $r'(\Ov)=(0)_{ij}$
implies that one can computably find an integer $k_2\geq k_1$
satisfying
$\|r'(\vec z)\|\leq c/2$
for all $|\vec z|\leq2^{-k_2}$.
Here, $\|B\|:=\sqrt{\sum_i\sum_j |b_{ij}|^2}$
denotes the \textsf{square sum norm} on matrices
which is known to be submultiplicative:
$|B\cdot\vec x|\leq\|B\|\cdot|\vec x|$.
Consequently,
by taking the norm on both sides of the \textsf{Mean Value Theorem}
$$  r(\vec y+\vec h)-r(\vec y)  \quad=\quad
       \bigg(\int_0^1 \big(r'(\vec y+t\vec h)\big)\,dt\bigg)
       \cdot \vec h \enspace , $$
it follows with $\vec h:=\vec x-\vec y$ 
that for all $\vec x,\vec y\in\cball(\Ov,2^{-k_2})$ we have:
\begin{equation} \label{eqMeanvalue}
|r(\vec x)-r(\vec y)| \;\leq\;
\bigg(\int_0^1 \| r'(\underbrace{\vec y+t\vec h}_{
  \in\cball(\Ov,2^{-k_2})\text{ convex}})\|\,dt\bigg)
\cdot|\vec h|
\;\leq\;\tfrac{c}{2}\cdot|\vec x-\vec y|
\enspace .
\end{equation}
\item
This asserts injectivity of $\tilde f\big|_{\cball(\Ov,2^{-k_2})}$.
Indeed, $\tilde f(\vec x)=\tilde f(\vec y)$ implies with
Equation~(\ref{eqDifferential}) that
$A\cdot\vec x+r(\vec x)=A\cdot\vec y+r(\vec y)$
and thus
$$ c\cdot|\vec x-\vec y| \;\leq\; |A\cdot(\vec y-\vec x)|
  \;=\; |r(\vec x)-r(\vec y)| \;\overset{(\ref{eqMeanvalue})}{\leq}\;
    \tfrac{c}{2}\cdot|\vec x-\vec y| \enspace : $$
a contradiction for $\vec x\not=\vec y$.
\end{itemize}
We may thus apply Lemma~\ref{lemDeimling} to obtain
some $\ell\in\IN$ such that
any $\vec y\in\ball\big(\tilde f(\Ov),2^{-\ell}\big)$ is
the image of one and exactly one
$\vec x\in\ball(\Ov,2^{-k_2})\subseteq\tilde U$.
Finally setting $\tilde g(\vec y):=\vec x$ shows that $\tilde f$
does have a local right inverse.
\qed\end{proof}

The first part of Theorem~\ref{thInverse}
claims the right inverse we have just constructed
to be computable and differentiable with computable derivative:
\begin{myclaim} \label{clInverseA}
Let $U\subseteq\IR^n$ be r.e. open,
$f:U\to\IR^m$ computable with computable derivative,
and $\vec x_0\in U$ computable such that $\rank f'(\vec x_0)=m$. 
Then there exists
a computable $C^1$ function $g$ with computable derivative
on some open ball $\ball\big(f(\vec x_0),\delta\big)\subseteq\IR^m$
satisfying (\ref{eqInverse}).
\end{myclaim}
\begin{proof}
Recall from the proof of Claim~\ref{clInverseB}
the reduction from the case $n\geq m$
to the case $n=m$ leading to a function $\tilde f$ instead of $f$
which turned out to be injective on some
$2^{-k'}$-ball around $\vec x_0$.
Let $\tilde U:=\ball(\vec x_0,2^{-k'-1})$
and apply to $\tilde f$ the classical Inverse Mapping Theorem,
in particular the last line of Fact~\ref{facInverse}:
It asserts $\tilde f$ to have on some (possibly smaller)
open ball
$\ball(\vec x_0,\tilde\varepsilon)\subseteq\ball(\vec x_0,2^{-k'-1})$
around $\vec x_0$ a unique and continuously differentiable
local inverse $\tilde g$. For any $\vec y$ from its domain
$\ball\big(\tilde f(\vec x_0),\tilde\delta\big)$,
the value $\tilde g(\vec y)$ is according to Equation~(\ref{eqInverse})
the unique
$\vec x\in\ball(\vec x_0,\tilde\varepsilon)$ with
$\tilde f(\vec x)=\vec y$.
Since $\tilde f$ is injective on
$\ball(\vec x_0,2^{-k'})
\supseteq\closure{U}\supseteq\ball(\vec x_0,\tilde\varepsilon)$,
$\tilde g(\vec y)$ is the unique zero of
$\vec x\mapsto f(\vec x)-\vec y$ on $\closure{U}$.
Computing $\vec y\mapsto\tilde g(\vec y)$ can thus
be performed by finding this zero by virtue of
Lemma~\ref{lemZero}; to actually apply it,
straight-forward scaling and translation effectively
reduces $\closure{U}=\cball(\vec x_0,2^{-k'-1})$ to
to $\cball(\Ov,1)$.

Differentiability of $\tilde g$ is asserted already classically.
Moreover, the \textsf{Chain Rule of Differentiation} yields
the formula $\tilde g'(\vec y)=\tilde f'\big(\tilde g(\vec y)\big)^{-1}$
which (\person{Cramer}'s Rule and computability of determinants)
reveals that $g'$ is computable as well.
\qed\end{proof}

\COMMENTED{
The following lemma, already announced and used above,
contains a sort of weak perturbation result on open functions.
\begin{lemma} \label{lemSurjective} 
Let $\varphi:\cball(\Ov,1)\subseteq\IR^n\to\IR^m$ satisfy $\varphi(\Ov)=\Ov$.
Furthermore let
$\psi:\cball(\Ov,2\delta)\subseteq\IR^m\to\cball(\Ov,1)\subseteq\IR^n$
be continuous such that
$\varphi\big(\psi(\vec y)\big)=\vec y$ for all
$\vec y\in\cball(\Ov,2\delta)$.
Consider finally continuous
$h:\cball(\Ov,1)\subseteq\IR^n\to\cball(\Ov,\delta)\subseteq\IR^m$.
Then the range of $\varphi+h$ covers $\cball(\Ov,\delta)$.
\end{lemma}
\begin{proof}
Let $\|\vec y\|\leq\delta$ be arbitrary. Then consider the function
$$  \cball(\Ov,1) \quad\ni\quad \vec x \quad\mapsto\quad
  \psi\big(\vec y - h(\vec x)\big) \enspace . $$
As $\|\vec y\|\leq\delta$ and $\|h(\vec x)\|\leq\delta$ on
$\cball(\Ov,1)\subseteq\IR^n$ and since furthermore
$\psi:\cball(\Ov,2\delta)\to\cball(\Ov,1)$, this function
maps back to the closed unit ball. A consequence to its continuity,
\person{Brouwer}'s \textsf{Fixed-Point Theorem} asserts the existence
of a fixed point: $\vec x=\psi\big(\vec y-h(\vec x)\big)$.
Applying $\varphi$ to both sides of this equality yields
$$ \varphi(\vec x) \quad=\quad \varphi\circ\psi\big(\vec y-h(\vec x)\big)
  \quad=\quad \vec y-h(\vec x) $$
by the right inverse property, and so $(\varphi+h)(\vec x)=\vec y$
as was intended to show.
\end{proof}
}
\section{Computable Open Semi-Algebraic Functions are Effectively
Open}
\label{secTarski}
Open Mapping Theorems give
conditions for continuous functions to be
open. However being only sufficient, they
miss many continuous open functions.
\begin{example} \label{exUnapplicable}
Let $f:\IR^3\to\IR^2$, $(x,y,z)\mapsto (x^3+z^2,y^3+z^2)$.
Then $f$ is open although no item from Fact~\ref{facOpen}
is applicable: a) fails due to the 2D range,
b) fails due to nonlinearity,
c) fails because $f'(\Ov)=\Ov$,
and d) fails as $f$ lacks injectivity everywhere.
\end{example}
Section~\ref{secEffectivized} of the present work
provided effectivizations of those classical results
where the prerequisites were strengthened from continuity
to computability in order to assert, in addition
to openness, \emph{effective} openness.
They therefore cannot be applied to cases such as
Example~\ref{exUnapplicable} where the
classical theorems fail already.
The main result of this section is of a different kind
in that it requires openness in order to conclude
effective openness. It is concerned with semi-algebraic
functions in the sense of, e.g., \mycite{Section~2.4.2}{Tarski}.
\begin{definition} \label{defSemiAlgebraic}
Let $F\subseteq\IR$ denote a field. ~
A set ~$S\subseteq\IR^n$~ is \emph{basic semi-algebraic over $F$} if
$$
S\quad=\quad
\big\{ \vec x\in\IR^n \;:\;
  p_1(\vec x)\geq0 \;\wedge\;\ldots\wedge\;
  p_k(\vec x)\geq0 \;\;\wedge\;\;
  q_1(\vec x)>0 \;\wedge\ldots\wedge\;
  q_{\ell}(\vec x)>0\big\}
$$
for certain $k,\ell\in\IN, \; p_1,\ldots,q_{\ell}\in F[X_1,\ldots,X_n]$,
that is, if $S$ is the set of solutions to some finite system of
polynomial inequalities both strict and non-strict with
coefficients from $F$. 
$S$ is \emph{semi-algebraic over $F$} if it is a
finite boolean combination (intersection and union)
of basic semi-algebraic sets over $F$. 
A partial function $f:\subseteq\IR^n\to\IR^m$ is
\emph{semi-algebraic over $F$} if
~$\Graph(f)=
\big\{(\vec x,\vec y):\vec x\in\dom(f)\wedge\vec y=f(\vec x)\big\}
\subseteq\IR^{n+m}$~ is semi-algebraic over $F$. 
In the case $F=\IR$, the indication ``\emph{over $F$}''
may be omitted.
\end{definition}

The class of semi-algebraic functions is very rich:
\begin{example} \label{exSemiAlgebraic}
\begin{enumerate}
\item[a)] Any rational function $f\in\IR(X_1,\ldots,X_n)$
 is semi-algebraic.
\item[b)] The roots of a univariate polynomial
  $p=\sum_{i=0}^{n-1} p_i\cdot x^i\in\IR[X]$,
  considered as a partial function of
  its coefficients $(p_0,\ldots,p_{n-1})$,
  are semi-algebraic.
\item[c)] For semi-algebraic $f$ and $g$, both composition
  $g\circ f$ and juxtaposition $(f,g)$ are again
  semi-algebraic.
  ~Projection ~$\IR^{n+m}\to\IR^n$, ~$(\vec x,\vec y)\mapsto\vec x$~
 is also semi-algebraic.
\end{enumerate}
\end{example}
\begin{proof}
a) ~
  Let $f=p/q$ with co-prime $p,q\in\IR[X_1,\ldots,X_n]$.
  Observe that
  $$(\vec x,y)\in\Graph(f) \quad\Leftrightarrow\quad
  q(\vec x)\not=0 \;\wedge\; p(\vec x)=y\cdot q(\vec x)$$
  which is a boolean combination of polynomial inequalities. ~
For b) and c) as well as for further
  examples of semi-algebraic functions,
  refer to \mycite{Section~2.4.2}{Tarski}.
\qed
\end{proof}

The main result of the present section
thus covers many more in addition to Example~\ref{exUnapplicable}.
\begin{theorem} \label{thTarski}
Let $f:\subseteq\IR^n\to\IR^m$ be
computable, open, and semi-algebraic over $\IRc$
with open $\dom(f)=:X$.
Then $X$ is r.e. and $f$ is effectively open.
\end{theorem}

\subsection{Applications of Quantifier Elimination to Recursive Analysis}
Quantifier elimination is an important tool in the algebraic framework
of computability and complexity \cite{BCSS,Tarski}. Its reliance on
(in)equality as a decidable primitive seemingly renders it useless
for the framework of Recursive Analysis.
It does however have interesting consequences
to \emph{non}-uniform computability as revealed in this section.

The following Lemma
will be applied to $E:=\IR$
and $F:=\IRc$ the set of computable real numbers,
a real closed field \mycite{Corollary~6.3.10}{Weihrauch},
but might be of independent interest and is therefore
formulated a bit more generally.
\begin{lemma} \label{lemPoly}
\begin{enumerate}
\item[a)] Let $F$ denote a real closed field with field extension $E$
  and $f\in F[X_1,..,X_n]$. If 
  ~$g\in E[X_1,\ldots,X_n]$~ divides $f$
  considered as polynomial
  over $E$, then $\lambda g\in F[X_1,\ldots,X_n]$~
  for some non-zero ~$\lambda\in E$.
\item[b)] Let $F$ denote a field with extension $E$.
  If $f,g\in F[X_1,\ldots,X_n]$ and
  $h\in E[X_1,\ldots,X_n]$ is a $\gcd$
  of $f$ and $g$ considered as polynomials over $E$,
  then $\lambda h\in F[X_1,\ldots,X_n]$ for some
  non-zero $\lambda\in E$.
\item[c)]
  Let $\emptyset\not=X\subseteq\IR^n$ be r.e. open.
  Suppose $p,q\in\IR[X_1,\ldots,X_n]$ are coprime with
  $q(\vec x)\not=0$ for all $\vec x\in X$
  and such that $p/q:X\to\IR$ is computable. 
  Then $\lambda p,\lambda q\in\IR_c[X_1,\ldots,X_n]$ for
  some non-zero $\lambda\in\IR$; that is,
  the coefficients of the rational function $p/q$
  may w.l.o.g. be presumed computable.
\vspace*{-0.3ex}\end{enumerate}
\end{lemma}
\begin{proof}
\begin{enumerate}
\item[b)]
  In the uni-variate case $n=1$, this follows from the
  \textsf{Euclidean Algorithm} since its calculation of the $\gcd$
  uses only arithmetic operations $+,-,\times,\div$ and thus
  remains within the coefficient field of the input polynomials
  $f$ and $g$. In the multi-variate case, the $\gcd$ is still
  well-defined (up to multiples $\lambda\in E$) based on
  unique factorization in $E[X_1,\ldots,X_n]$
  \mycite{Exercise~4.\S2.9}{Ideals}. Moreover, it can be
  calculated via \textsf{Gr\"{o}bner Bases}
  \mycite{Propositions~4.\S3.13+14}{Ideals}, again
  using only arithmetic operations and thus remaining
  within the field $F$.
\item[a)] The equation $f=g\cdot h$ of $n$-variate polynomials over $E$
  translates to a finite bilinear system of (in-)equalities for the
  $\calO(d^n)$ coefficients of $g$ and those of $h$ with coefficients
  from $F$, $d:=\deg(g)$.
  The absolute terms for instance must satisfy $g_0\cdot h_0=f_0$
  and the leading term must be non-zero.
  By the \textsf{Tarski-Seidenberg Transfer Principle}
  \mycite{Theorem~2.78}{Tarski}, this system is solvable over
  $E$ (with solution $g|f$) iff it is solvable over $F$
  (with solution $\tilde g|f$).
  The condition on the leading term asserts $\deg(g)=\deg(\tilde g)$.
  Again by uniqueness of factorization it follows that
  $\tilde g=\lambda g$ for some non-zero $\lambda\in E$.
\item[c)]
  Let $d>\deg(p)+\deg(q)$ and consider a
  $(d\times d\times\ldots\times d)$--grid
  of computable vectors $\vec x\in X$, that is,
  $n$ sets $X_1,\ldots,X_n\subseteq\IRc$ of
  cardinality $|X_i|=d$ such that
  $X_1\times X_2\times\ldots\times X_n\subseteq X$;
  such exist because $\IRc$ is dense and $X$ is non-empty and open.
  By prerequisite, $y_j:=p(\vec x_j)/q(\vec x_j)\in\IRc$
  for each $\vec x_j\in\prod_i X_i$, $j=1,\ldots,d^n$.
  Expanding the equations
  ~$p(\vec x_j)-y_j\cdot q(\vec x_j)=0$~
  in the multinomial standard basis yields a homogeneous system of linear
  equations with respect to
  the coefficients of both $p$ and $q$ to be solved for.

  On the other hand the system itself is composed from
  (products of components of) computable reals $\vec x_j$ and $y_j$.
  It follows from \mycite{Corollary~15}{LA} that
  this system also admits a \emph{computable} non-zero
  solution $\tilde p,\tilde q\in\IRc[X_1,\ldots,X_n]$.
  In particular, $\tilde p/\tilde q$ is defined and coincides
  with $p/q$ almost everywhere on $X$.

  For $h:=\gcd(\tilde p,\tilde q)$,
  $\hat p:=\tilde p/h$ and $\hat q:=\tilde q/h$
  are coprime and, based on Items a) and b),
  still belong to $\IRc[X_1,\ldots,X_n]$.
  Moreover it holds that $\hat p\cdot q=\hat q\cdot p$
  by uniqueness of multivariate polynomials on the
  grid $X_1\times\ldots X_n$
  (e.g. \textsf{Schwartz-Zippel Lemma}). As
  $q$ divides $\hat p\cdot q=\hat q\cdot p$,
  coprimality with $p$ requires it to divide $\hat q$.
  Similarly $\hat q$ divides $q$. Thus
  $\hat q=\lambda q$ for some non-zero $\lambda\in E$
  and consequently $\hat p=\lambda p$.
\qed
\end{enumerate}
\end{proof}

It is well-known in Recursive Analysis
that equality of reals
lacks even semi-decidability.
Surprisingly it becomes decidable
for rational arguments to real polynomial
equations:
\begin{proposition} \label{proTarski}
\begin{enumerate}
\item[a)]
  Let $p\in\IR[X_1,\ldots,X_n]$ denote an $n$-variate polynomial.
  \\ Then ~$\{\vec x\in\IQ^n:p(\vec x)=0\}$~ is decidable
  in the classical (i.e., Type-1) sense.%
\item[b)]
  Let $\Psi(X_1,\ldots,X_n)$ denote a finite Boolean combination of
  polynomial equalities and inequalities in variables $X_1,\ldots,X_n$
  with computable real coefficients.
  \\ Then ~$\{\vec x\in\IQ^n:\Psi(\vec x)\}$~ is 
  (classically) semi-decidable.
\item[c)]
  Let $X\subseteq\IR^n$ be open and semi-algebraic over $\IRc$.
  Then $X$ is r.e, that is, $\thetaL{n}$--computable.
\end{enumerate}
\end{proposition}
\pagebreak[2]
\begin{proof}
\begin{enumerate}
\item[b)]
  Without loss of generality, $\Psi$ consists --- apart from equalities ---
  of \emph{strict} inequalities only; otherwise replace
  any ``$p(\vec x)\leq0$''
  with ``$p(\vec x)<0\;\vee\;p(\vec x)=0$''.
  Since $p\in\IRc[X_1,\ldots,X_n]$ is computable by assumption,
  strict inequalities are obviously semi-decidable;
  and equalities are even decidable by virtue of
\item[a)]
  Let
  $$ p(\vec x)
    = \sum_{k_1=0}^{d_1} \ldots \sum_{k_n=0}^{d_n}
      a_{(k_1,\ldots,k_M)} \cdot x_1^{k_1}\cdots x_n^{k_n} $$
  with $a_{(k_1,\ldots,k_M)}\in\IRc$.
  Choose\footnote{and observe the strong non-uniformity
  inherent in this step; for example a still open problem
  of number theory asks whether $e\cdot\pi$ or
  $e+\pi$ is rational \mycite{p.153}{Numbers}.}
  among these $a_{\vec k}$
  a basis $\{b_0=1,b_1,\ldots,b_m\}$ for the
  finite-dimensional $\IQ$--vector space
  $V:=\big\{q_0+
    q_{\vec k}a_{\vec k}+\ldots+q_{(d_1,\ldots,d_n)}a_{(d_1,\ldots,d_n)}
    :q_{\vec k}\in\IQ\big\}$.
  Consequently, each coefficient of $p$ is of the form
  ~$a_{\vec k} = \sum_{i=0}^m A_{i,\vec k} b_i$~ with
  fixed $A_{i,\vec k}\in\IQ$.
  Now for given $\vec x\in\IQ^n$,
  $$ 0 \quad=\quad p(\vec x) \quad=\quad
    \sum_{i=0}^m b_i \cdot \underbrace{\sum_{k_1=0}^{d_1}
     \ldots \sum_{k_n=0}^{d_n} A_{i,\vec k}\cdot
       x_1^{k_1}\cdots x_n^{k_n}}_{=:R_i(\vec x)\in\IQ} $$
  holds if and only if $R_i(\vec x)=0$ for all $i=0,\ldots,m$
  because the $b_i$ are linearly independent over $\IQ$.
  The equalities ~$R_i(\vec x)=0$~ in turn are of course
  decidable by means of exact rational arithmetic.
\item[c)]
  Let $\vec x\in\IQ^n$ and $0<r\in\IQ$. Then
  ~``$\ball(\vec x,r)\subseteq X$''~ is equivalent to
  $$\forall \vec y\in\IR^n: \quad
  \Big( \sum_{i=1}^n (y_i-x_i)^2<r^2 \;\Rightarrow\;
   \vec y\in X\Big)$$
  a first-order formula $\Phi(\vec x,r)$ in the language of ordered fields
  with coefficients by assumption from the real closed field $\IRc$.
  By \person{Tarski}'s \textsf{Quantifier Elimination}\footnote{This
  proof bears some similarity to \cite{Equality};
  there however the sets under consideration are BSS-semi-decidable
  (i.e., roughly speaking, countable unions of semi-algebraic ones)
  and therefore $\thetal$--computable (recursively enumerable) 
  only \emph{relative} to the Halting problem.},
  there exists an equivalent quantifier-free formula
  $\Psi(\vec x,r)$ over $\IRc$ \mycite{Theorem~2.74}{Tarski};
  but for rational $(\vec x,r)$, ~$\Psi(\vec x,r)$ is
  semi-decidable according to b)
  and 
  ~$X\;=\;\bigcup\big\{\ball(\vec x,r) :
  \vec x\in\IQ^n, 0<r\in\IQ, \ball(\vec x,r)\subseteq X\big\}$~
  is therefore $\thetaL{n}$--computable.
\qed
\end{enumerate}
\end{proof}

\subsection{Proof of Theorem~\ref{thTarski} and Consequences}

\begin{proof}[Theorem~\ref{thTarski}]
The domain of $f$ is semi-algebraic over $\IRc$
according to \mycite{Proposition~2.81}{Tarski}
and thus r.e. due to Proposition~\ref{proTarski}c).
Similarly to the proof there, we observe:
\begin{multline*}
\cball\big(f(\vec x),s\big)
\subseteq f\big[\ball(\vec x,2^{-k})\cap X\big]
\qquad\Longleftrightarrow \\[1ex]
\exists\vec v\in\IR^n\:
\forall\vec y\in\IR^m\:
\exists\vec u\in\IR^m:\quad
(\vec x,\vec v)\in\Graph(f)
\;\wedge\;
(\vec u,\vec y)\in\Graph(f)
\;\;\wedge\quad \\
\Big( \sum_{j=1}^m (y_j-v_j)^2\leq s^2
\;\Rightarrow\; \sum_{i=1}^n (x_i-u_i)^2<2^{-2k}
\Big)
\end{multline*}
Since the latter is a first-order formula $\Phi(\vec x,s)$,
by assumption with coefficients from $\IRc$,
there exists by \mycite{Theorem~2.74}{Tarski}
an equivalent quantifier-free formula $\Psi(\vec x,s)$
again over $\IRc$.
This in turn is semi-decidable
for rational $(\vec x,s)$ by virtue of
Proposition~\ref{proTarski}b)
so that $\Moo_f$ can be approximated from
below on $\IQ^n$.
Now apply the Lemma below.
\qed
\end{proof}

Example~\ref{exDense} illustrated that in
Lemma~\ref{lemModulus}b) as well a
Theorem~\ref{thModulus}b), it does not suffice
to consider $\Moo_f$ only on a dense subset of $X$.
On the other hand if $f$ is already asserted as
open, then computability of $\Moo_f$ on rationals
already does guarantee effective openness:
\begin{lemma} \label{lemDense}
Let $X\subseteq\IR^n$ be r.e. open;
furthermore let $f:X\to\IR^m$ be computable and open.
If $\Moo_f:(X\cap\IQ^n)\times\IN\to\IR$ is
$(\nu^n\times\nu\to\myrhol)$--computable,
then $f$ is effectively open.
\end{lemma}
\begin{proof}
The goal is to
$\thetaL{m}$--compute $f[U]$, given a
$\thetaL{n}$--name of some open $U\subseteq X$.
To this end observe that, similarly to the proof of Claim~\ref{clTwo},
\begin{multline*}
f[U] \;\overset{\surd}{=}\;
 \bigcup_{\vec x\in U}
  \ball\Big(f(\vec x),\Moo_f\big(\vec x,G(U,\vec x)\big)/2\Big) \\[-1ex]
\supseteq\quad
 \bigcup_{\vec x'\in U\cap\IQ^n}
  \ball\Big(f(\vec x'),\Moo_f\big(\vec x',G(U,\vec x')\big)/2\Big)
\;=:\; V
\end{multline*}
with $G$ from Lemma~\ref{lemMulti}a).
And, again, the countable union
$V$ can be $\thetaL{m}$-computed because
$\IQ^n\times\IR\ni(\vec z,r)\mapsto
  \ball\big(f(\vec z),r\big)$
is $(\nu^n\times\myrhol\to\thetaL{m})$--computable
and the multi-valued mapping
$h:\IQ^n\ni\vec z\mapstoto\Moo_f\big(\vec z,G(U,\vec z)\big)/2$
is $(\nu^n\toto\myrhol)$--computable by assumption.
It remains to show that, again,
the reverse inclusion ``$f[U]\subseteq V$'' holds as well
for a suitable computable subfunction. More precisely
w.l.o.g. replace $G$ from Lemma~\ref{lemMulti}a)
by $\tilde G$ according to Lemma~\ref{lemMulti}b)
such that $G(U,\cdot)$ is bounded on compact subsets of $\IR^n$.
Now consider some $\vec x\in U\setminus\IQ^n$. We show that then
$f(\vec x)\in V$:

Let some compact ball $\cball:=\cball(\vec x,r)$
be contained in $U$
and take an upper bound $L\in\IN$ for $\tilde G(U,\cdot)$ on $\cball$.
By assumption, $\delta:=\Moo_f(\vec x,L+1)$ is strictly positive.
The computable $f$ is continuous so that,
for some $0<r'\leq\min\{r,2^{-L-1}\}$,
$f(\vec x')\in\ball\big(f(\vec x),\delta/2\big)$
whenever $\vec x'\in\cball':=\cball(\vec x,r')$.
$\IQ^n$ being dense in $U$, there exists some
rational $\vec x'\in\cball'$.
Now observe that
\vspace*{-0.5ex}
\begin{itemize}
\item[i)]
  $\cball(\vec x,2^{-(L+1)})\subseteq\cball(\vec x',2^{-L})$
  by choice of $\vec x'$, thus 
  $f\big[\,\cball(\vec x,2^{-(L+1)})\big]\subseteq
   f\big[\,\cball(\vec x',2^{-L})\big]$;
\item[ii)]
  from 
  continuity of $f$ it follows
  $\cball\big(f(\vec x'),s-\tfrac{\delta}{2}\big)
  \subseteq\cball\big(f(\vec x),s\big)$
  for any $s\geq\tfrac{\delta}{2}$.
\item[iii)]
  Combining i) and ii) yields
  $\Moo_f(\vec x',L)+\tfrac{\delta}{2}\geq\Moo_f(\vec x,L+1)=\delta$
  because, by Equation~(\ref{eqMoo1}), 
  $\Moo_f(\vec x',L)$ is the
  \emph{supremum} of feasible radii $s$.
\item[iv)]
  Any $\ell'\in\tilde G(U,\vec x')$ has $\ell'\leq L$ by choice of $L$;
  ~ therefore
\item[v)]
  $\Moo_f(\vec x',\ell')\geq\Moo_f(\vec x',L)$
  as $\Moo_f(\vec x',\cdot)$ is monotonic according
  in Equation~(\ref{eqMoo1}).
\vspace*{-0.5ex}
\end{itemize}
We conclude that $\delta':=\Moo_f(\vec x',\ell')\geq\tfrac{\delta}{2}$
and $f(\vec x)\in\ball\big(f(\vec x'),\delta'\big)\subseteq V$.
\qed
\end{proof}

\COMMENTED{
  Observe that $f$ is computable in the sense of Recursive Analysis.
  Fix $\delta>0$. For $\vec y\in\ball(0,\delta)$,
  $\vec y\in f\big[\cball(0,1)]$ iff
  the closed set $B:=f^{-1}[\{\vec y\}]$ intersects $\cball(0,1)$.
  Both being $\psiG{n}$--computable, so is their intersection:
  \textsc{Theorems~6.2.4.2} and \textsc{5.1.13.2} in \cite{Weihrauch}.
  Whether this intersection is empty can be semi-decided according
  to Lemma~\ref{lemCover}.
  Trying \emph{all} rational $\vec y\in\ball(0,\delta)$
  thus allows to semi-decide whether
  $\ball(0,\delta)\not\subseteq f\big[\ball(0,1)]$.
  Doing so dove-tailed for all rational $\delta>0$
  leads to upper approximations for $\delta_0$.
}

\begin{corollary}
If the rational functions
$f_i\in\IR(X_1,\ldots,X_m)$
are computable for $i=1,\ldots,n$
and the function
$(f_1,\ldots,f_m):\subseteq\IR^n\to\IR^m$
is open, then it is effectively open.
\end{corollary}
\begin{proof}
According to Example~\ref{exSemiAlgebraic}a),
$f_i$ as well as its domain is semi-algebraic;
in fact semi-algebraic {over $\IRc$} 
by virtue of Lemma~\ref{lemPoly}c).
Now apply Theorem~\ref{thTarski}.
\end{proof}

In Theorem~\ref{thTarski}, $f$ was explicitly required
to be semi-algebraic over $\IRc$; yet it seems reasonable,
similarly to Lemma~\ref{lemPoly}c), to
\begin{conjecture}
Let $F\subseteq\IR$ be a real closed subfield.
Furthermore let $f:\subseteq\IR^n\to\IR$ be
continuous and semi-algebraic (over $\IR$!) with
$\dom(f)$ semi-algebraic over $F$ and such that
$f(\vec x)\in F$ whenever $\vec x\in F^n\cap\dom(f)$.
Then $f$ is semi-algebraic already over $F$.
\end{conjecture}
\section{Effective Openness and Computability} \label{secConverse}
The preceding sections presented sufficient conditions
for a computable function $f$ to be effectively open.
The present one aims more generally at the logical
relation between openness, continuity, effective openness,
and computability of real functions.

The two classical properties for instance are well-known
mutually independent: Continuity does not imply openness;
nor does openness require continuity.
(Counter-)Examples~c) and d) below
reveal that the same still holds under effectivized prerequisites.
\begin{example} \label{exConverse} ~\vspace*{-1ex}
\begin{enumerate}
\item
 There exists a function $h:\IR\to\IR$ such that $h[(a,b)]=(0,1)$ for any
 $a<b$.
\item
 There exists an open but not effectively open real function.
\item
 There exists a computable but not open real function;
  \quad e.g. ~$f:\IR\to\IR$, $x\mapsto 0$.
\item
 There exists an effectively open but uncomputable real function.
\item
 There exist open functions $f_1,f_2:\IR\to\IR$ such that
 $f_1\not=f_2$ but
 $f_1[U]=f_2[U]$ for any open $U\subseteq\IR$.
\end{enumerate}
\end{example}
\begin{proof}
\begin{enumerate}
\item Cf., e.g., item no.100 in the \textsc{Guide}
  preceding \cite{Counterexamples}.
\item Let $u$ be right-uncomputable and $v>u$ be left-uncomputable.
  Let $g:\IR\to\IR$, $g(x)=u+(v-u)x$.
  Then, with $h$ from a), $g\circ h:\IR\to\IR$
  has image $(u,v)$ for any non-empty open $U$
  and is thus open; but under
  $\thetal$--computable $U:=(0,1)$,
  this image lacking $\thetal$-computability
  \cite[\textsc{Example~5.1.17.2}a)]{Weihrauch}
  reveals that $g\circ h$ is not effectively open.
\stepcounter{enumi}
\item The function $h$ from a) is open but
 maps the compact interval $[0,1]\subseteq\IR$ to the non-compact
 interval $(0,1)$ 
 $$   (0,1) \quad=\quad h\big[(\tfrac{1}{3},\tfrac{2}{3})\big]
   \quad\subseteq\quad h\big[[0,1]\big] 
   \quad\subseteq\quad h\big[(-1,2)\big] \quad=\quad (0,1) $$
 and thus cannot be continuous nor computable.
 For $(\thetal\to\thetal)$-computing $U\mapsto h[U]$,
 it suffices to output a $\thetal$-name of $(0,1)$
 \cite[\textsc{Example~5.1.17.2}c)]{Weihrauch}
 independent of the input $U\not=\emptyset$.
 The test ``$U\not=\emptyset$" is obviously semi-decidable, 
 formally: $\Open\setminus\{\emptyset\}$ is $\thetal$-r.e. 
 \mycite{Definition~3.1.3.2}{Weihrauch}.
\item
 Let $f_1:=h$ from a) and $f_2:=g\circ h$ with
 $g:\IR\to\IR$, $g(y)=y^3$, open as composition of two open functions.
 As $h[(0,1)]=(0,1)$, there is some $x\in(0,1)$ such that
 $h(x)=y:=\tfrac{1}{2}$. Then $f_2(x)=\tfrac{1}{8}$ reveals
 that $f_1\not=f_2$.
\qed
\end{enumerate}
\end{proof}

Attempts to strengthen Examples~\ref{exConverse}c) and d)
immediately raise the following
\begin{question} \label{quConjecture}
\begin{enumerate}
\item
  Is there a computable, open but not effectively open real function?
\item
  Is there a continuous, effectively open but uncomputable real function?
\end{enumerate}
\end{question}

Regarding Theorem~\ref{thOpen}a), a putative example for
Question~\ref{quConjecture}a)
must have domain and range both of dimension at least two,
that is, a graph living in $\IR^d$ for some $d\geq4$.
Moreover its graph cannot be semi-algebraic because
of Theorem~\ref{thTarski}.
Concerning candidates to \ref{quConjecture}b),
the following result allows to restrict research
to functions with one-dimensional range
on domains of dimension at least two.
\begin{theorem} \label{thConverse} ~\vspace*{-1ex}
\begin{enumerate}
\item
On r.e. open $X\subseteq\IR$,
any continuous and effectively open
$f:X\to\IR$~ is also computable.
\item
Let $X\subseteq\IR^n$ be r.e. open,
$f=(f_1,\ldots,f_m):X\to\IR^m$ continuous and effectively open
but not computable. \\
Then some $f_i:X\to\IR$, too,
is continuous and effectively open, but not computable.
\end{enumerate}
\end{theorem}
\begin{proof}
\begin{enumerate}
\item[b)]
Recall that the projections
$\proj_i:\IR^n\to\IR$, $(x_1,\ldots,x_n)\mapsto x_i$
are computable (hence continuous) and open; even effectively open:
Theorem~\ref{thOpen}a) or b).
By closure under composition, the component functions
$f_i=\proj_i\circ f:X\to\IR$ 
are therefore continuous and effectively open themselves.
Regarding that a vector-valued $f$ is computable iff its components are
\mycite{Lemma~4.1.19.5}{Weihrauch},
it follows that at least some $f_i$ cannot be computable.
\item[a)]
To evaluate $f$ at a given $x\in X$,
we are given two monotonic sequences $(u_j)_{_j}$ and $(v_j)_{_j}$
of rational numbers
converging to $x$ from below and above, respectively.
As $x\in X$ is open, the entire interval
$[u_{J},v_{J}]$ belongs to $X$ for some $J\in\IN$;
and, since $(u_j)$ and $(v_j)$ are respectively increasing and decreasing
to $x$, also $x\in [u_j,v_j]\subseteq X$ for all $j\geq J$.
In fact, such $J$ can be found effectively because
the property ``$[u_J,v_J]\subseteq X$'' is semi-decidable
by virtue of \cite[\textsc{Lemma~4.1}b)]{MLQ2}.
\\
Now, for each $j\geq J$,
$\thetal$-compute the open intervals
$U_j:=(u_j,u_{j+1})$ and $V_j:=(v_{j+1},v_j)$
as well as
(by prerequisite) their images $f[U_j]$ and $f[V_j]$
and choose rational numbers $a_j\in f[U_j]$ and $b_j\in f[V_j]$.
According to Lemma~\ref{lemMonoton} below,
both sequences $(a_j)$ and $(b_j)$
converge to $f(x)$ monotonically from different sides;
and comparing $a_J$ to $b_J$ immediately reveals which
one constitutes the lower and which one the upper approximations.
\qed
\end{enumerate}
\end{proof}

The following lemma can be regarded as a one-dimensional converse
to Fact~\ref{facOpen}d) because it implies that, for arbitrary
open $X\subseteq\IR$, a continuous open function $f:X\to\IR$
is injective on any connected component of $f$.
\begin{lemma} \label{lemMonoton}
Let $X\subseteq\IR$ be open and connected,
$f:X\to\IR$ continuous and open. \\
Then $f$ is either strictly
increasing or strictly decreasing.
\end{lemma}
\begin{proof}
Let $a,b,c\in X$, $a<b<c$. W.l.o.g. presuming $f(a)<f(c)$ --- otherwise
consider the negative inverse of $f$ --- we show $f(a)<f(b)<f(c)$.
Now suppose for instance that $f(b)>f(c)$.
As $f$ is continuous on $[a,c]$, it attains its maximum therein
at some $x\in[a,c]$ with a value $f(x)\geq f(b)>\max\{f(a),f(c)\}$;
in particular, $x\in(a,c)$. Therefore, the interval $f[(a,c)]$ is
closed on its upper end contradicting that $f$ is open. 
By considering the minimum of $f$ on $[a,c]$,
the case $f(b)<f(a)$ similarly raises a contradiction.
\qed\end{proof}

\section{Conclusion} \label{secConclusion}
The present work investigated conditions for an open function
$f:\IR^n\to\IR^m$ to be \emph{effectively} open in the sense that
the image mapping $U\mapsto f[U]$
is $(\thetaL{n}\to\thetaL{m})$--computable.
This property is so to speak dual to function \emph{computability}
because the latter holds for $f:\IR^n\to\IR^m$
iff the \emph{pre}-image mapping $V\mapsto f^{-1}[V]$
is $(\thetaL{m}\to\thetaL{n})$--computable.
\begin{remark}
This characterization of computable real functions
gave in \textsc{Definition~6.1.6} of \cite{Weihrauch} rise
to a natural representation --- equivalent to many other ones
\cite[\textsc{Lemmas~6.1.7} and \textsc{6.1.10}]{Weihrauch} ---
for the space $C(X,\IR^m)$ of \emph{all} (not necessarily computable)
continuous functions $f:X\to\IR^m$, namely by $\thetaL{n}$-encoding,
for each open rational ball $B\subseteq\IR^m$,
the open set $f^{-1}[B]$.

Analogy might suggest to represent the family of \emph{all}
(not necessarily computable) open functions $f:X\to\IR^m$
by $\thetaL{m}$-encoding, for each open rational ball $B\subseteq\IR^n$,
the open set $f[B]$. However Example~\ref{exConverse}e)
reveals that such a representation would not be well-defined.
\end{remark}

It was already mentioned that
effectively open functions have applications in computations
on regular sets such as in solid modeling. For instance when
encoding bounded regular $R\subseteq\IR^d$
as a list of open rational balls with union dense in $R$
(representation $\cthetadl$),
this will render not only union and intersection computable but
also pre-image and 
image $R\mapsto g[R]$ under computable effectively open
functions $g$ \mycite{Theorem~3.9}{MLQ2}.
According to Theorem~\ref{thOpen}c), that requirement on $g$
is satisfied by any computably differentiable function with
regular derivative everywhere. However some $g$ might
be computably differentiable and open with $g'(\vec x)$
occasionally singular. The following result based on
\textsf{Sard's Theorem} 
asserts that,
even then, $R\mapsto g[R]$ is $\cthetadl$--computable:
\begin{theorem}
Let $g:\IR^n\to\IR^m$ be computable, open, and $C^1$ with computable
derivative $g'$. Then its image mapping on bounded regular sets
~$R\mapsto g[R]$~ is $(\cthetaL{n}\to\cthetaL{m})$--computable.
\end{theorem}
\begin{proof}
Let ~$U_0:=\displaystyle \big\{ \vec x\in\IR^n : \rank\big(g'(\vec x)\big)=m
\big\}$~ denote the set of regular points of $g$.
\\
Consider
the function $G:=\rank\circ\,g':\IR^n\to\IN$,
$\vec x\mapsto\rank\big(g'(\vec x)\big)$.
Because of its discrete range,
the pre-image $G^{-1}\big[(m-\tfrac{1}{2},\infty)\big]$
obviously coincides with $U_0$.
Moreover, being the composition of the lower semi-computable
$\rank$-function 
--- see \mycite{Proposition~6}{Linear} or
\cite[\textsc{Theorem~7}(i)]{LA} ---
with computable $g'$, $G$ is in particular lower semi-continuous
and $U_0\subseteq\IR^n$ therefore open
\mycite{Def.~2.8}{Rudin};
in fact r.e. open, see Lemma~\ref{lemSemipreimage} below.
\\
By Theorem~\ref{thOpen}c), at least the
the restriction $g\big|_{U_0}$ is thus effectively open.
So given as $\cthetaL{n}$-name for $R\subseteq\IR^n$
a $\thetaL{n}$-name for open $U\subseteq\IR^n$ with
$\closure{U}=R$, 
$\thetaL{n}$--compute $U\cap U_0$ according to
\mycite{Corollary~5.1.18.1}{Weihrauch};
then exploit effective openness of $g\big|_{U_0}$ to
$\thetaL{m}$--compute $V:=g[U\cap U_0]$.
\\
We claim that this yields a valid $\cthetaL{m}$-name
for the regular set $g[R]$, i.e.,
it holds that $\closure{V}=g[R]$. To this end,
observe that $\closure{U_0}=\IR^n$;
for if $A_0:=\IR^n\setminus U_0$ had non-empty
interior, then the set $V_0:=g\big[\interior{A_0}\big]$ of critical
values would be open (since $g$ is open by prerequisite)
and non-empty rather than having measure zero
as \textsf{Sard's Theorem} asserts \cite{Sard}.
$U_0$ thus being dense,
\cite[\textsc{Lemma~4.3}c) and \textsc{Lemma~4.4}d)]{MLQ2}
imply $\closure{U\cap U_0}=R$ and
$\closure{V}=g[R]$.
\qed\end{proof}

\begin{lemma} \label{lemSemipreimage}
Let $h:\IR^d\to\IR$ be lower semi-computable.
\\
Then the mapping
\quad $\IR \;\ni\; \alpha \;\mapsto\;
  h^{-1}\big[(\alpha,\infty)\big] \;\subseteq\;\IR^d $ \quad
is $(\myrhog\to\thetadl)$--computable.
\end{lemma}
\begin{proof}
Recall that lower semi-computability of $h$ means
that evaluation of $h$ at some $\vec x\in\IR^d$,
given open rational balls $\ball_j\ni\vec x$ of radius $r_j\to0$,
yields rational numbers $\beta_j$ tending from below
to $h(\vec x)$.
\\
So feed into this $h$-oracle \emph{all} open rational balls
$\ball_j\subseteq\IR^d$ and, whenever the answer $\beta_j$ is
strictly greater than $\alpha$ (semi-decidable,
given a $\myrhog$-name for $\alpha$),
report this $\ball_j$. The resulting sequence
obviously covers exactly ~$h^{-1}\big[(\alpha,\infty)\big]$
and consequently is a $\thetadl$-name for this set.
\qed\end{proof}

\paragraph{Acknowledgments:}
The author is grateful to the Danish
\textsf{Statens Naturvidenskabelige Forskningsr{\aa}d} (SNF)
for funding under project 21-04-0303 and to
\textsc{Klaus Meer} for the acquisition of that project.
Section~\ref{secTarski} of the present work
in fact emerged from his advanced lecture \textsf{DM82}
held 2005 in Odense.
Further gratitude is owed to an anonymous referee for
a careful review revealing an error in a previous
version of Lemma~\ref{lemMulti}.

\end{document}